\documentstyle[12pt,fleqn]{article}

\topmargin - 0.8cm

\renewcommand{\theequation}{\arabic{section}.\arabic{equation}}
\renewcommand{\thesection}{\arabic{section}.}

\catcode`@=11
\@addtoreset{equation}{section}
\catcode`@=12
\mathchardef\SGamma="7100

\begin{document}
\title{\vskip-1.7cm \bf  On covariant long-distance
modifications of Einstein theory and strong coupling problem}
\date{}
\author{A.O.Barvinsky}
\maketitle
\hspace{-8mm}
{\em Theory Department, Lebedev Physics
Institute, Leninsky Prospect 53, Moscow 117924, Russia}

\begin{abstract}
We present generic form of the covariant nonlocal action for
infrared modifications of Einstein theory recently suggested
within the weak-field curvature expansion. In the lowest order it
is determined by two nonlocal operators --- kernels of Ricci
tensor and scalar quadratic forms. In models with a low
strong-coupling scale this action also incorporates the strongly
coupled mode which cannot be perturbatively integrated out in
terms of the metric field. This mode enters the action as a
Lagrange multiplier for the constraint on metric variables which
reduces to the curvature scalar and enforces the latter to vanish
on shell. Generic structure of the action is demonstrated on the
examples of the Fierz-Pauli and Dvali-Gabadadze-Porrati models and
their extensions. The gauge-dependence status of the
strong-coupling and VDVZ problems is briefly discussed along with
the manifestly gauge-invariant formalism of handling the braneworld
gravitational models at classical and quantum levels.
\end{abstract}

\section{Introduction: new mechanism of small cosmological constant}

As is well known, the essence of the cosmological constant problem
(CCP) consists in the enormous gap between the average density of
energy in the modern Universe ${\cal E}\sim 10^{-29}
\mbox{g/cm}^3\sim (10^{-5} \mbox{eV})^4$, generating according to
Einstein equations the cosmological acceleration with the
Hubble constant $H_0$,
    \begin{eqnarray}
    H_0^2\sim G\,{\cal E},  \label{9.1}
    \end{eqnarray}
and the vacuum energy scale of all field theory models ranging
from electroweak theory, ${\cal E}\sim (1 \mbox{TeV})^4$, to
quantum gravity, ${\cal E}\sim (10^{19}$ \mbox{GeV})$^4$.
Old attempts to resolve this problem were based on building
the models with zero vacuum energy, incorporating supersymmetry which
protects the cosmological constant from renormalization. The
cosmological constant mechanism in such models is based on
cancellation of contributions of particles and their superpatners
and stops working in the phase with spontaneously broken
supersymmetry \cite{Weinberg}. Moreover, it becomes completely
meaningless in the framework of the cosmological acceleration
phenomenon \cite{accel1,accel2,accel3} for which the magnitude of
the cosmological constant, though being very small, is still
different from zero.

An alternative solution of CCP may be attributed to the curvature
scalar sector of the Einstein-Hilbert action, rather than to its
cosmological constant term. The smallness of $H_0^2$ in
(\ref{9.1}) can be explained not by small value of ${\cal E}$, but
by the smallness of the proportionality coefficient ---
gravitational coupling constant $G$. In other words, the
cosmological acceleration is so small not because the vacuum
energy is small, but rather because the latter is gravitating very
weakly. Special property of the vacuum energy as compared to other
local sources of the gravitational field is the degree of its
spacetime homogeneity. It is assumed that it is not clustering and
practically homogeneous at the horizon scale,
    \begin{eqnarray}
    \frac{\nabla\cal E}{\cal E}\sim H_0,  \label{9.2}
    \end{eqnarray}
and gravitates with its long-wavelength gravitational constant
$G_{LD}\ll G_P$ which is much smaller than the Planckian constant
determining the everyday physics in the range of galaxies,
planetary and solar systems, submillimeter Kavendish type
experiments \cite{Gundlach}, etc., $H_0^2\sim G_{LD}{\cal E}\ll
G_P{\cal E}$.

This idea, that was apparently formulated in such explicit form
for the first time in \cite{A-HDDG}, represents the replacement of
the fundamental gravitational constant in Einstein equations by
the nonlocal operator, which for reasons of covariance can be
regarded as a function of the d'Alembertian and which interpolates
between the Planckian value of the gravitational constant and its
long-distance magnitude\footnote{The idea of scale dependent
Newton constant was also put forward in \cite{ParSol}, though it
was not formulated in terms of the nonlocal operator.}
    \begin{eqnarray}
    G\,\Rightarrow\, G(\Box),\,\,\,\,
    G_P>G(\Box)>G_{LD}.      \label{9.3}
    \end{eqnarray}

Note that this mechanism of scale dependent gravitational constant
is not unique. The notion of scale includes not only the degree of
spacetime inhomogeneity, but also the field amplitude. Therefore,
the infrared modification of the theory can also be based on the
gravitational "constant" locally depending on some distinguished
physical fields --- sort of quintessence
\cite{Weinberg,quintessence}.  However, such modifications are
less universal because they are attributed to the behavior of
concrete quintessence field, while the mechanism of the nonlocal
replacement (\ref{9.3}) leads to modified Einstein equations
    \begin{eqnarray}
    R_{\mu\nu}-\frac12
    g_{\mu\nu}R=8\pi G(\Box)\,T_{\mu\nu},  \label{9.4}
    \end{eqnarray}
in which independently of the field content of the matter source
$T_{\mu\nu}=T_{\mu\nu}(x)$ the gravitational strength of the
latter is determined solely by the degree of its inhomogeneity in
$x$.

One of the first problems associated with (\ref{9.4}) is the lack
of general covariance --- no covariant action can generate such
equations of motion with a nontrivial $G(\Box)$. The solution of
this problem was suggested in \cite{nonloc} by viewing (\ref{9.4})
only as a first, linear in the curvature, approximation for
correct equations of motion, their covariant action being
constructed as a weak field expansion in powers of the curvature
with nonlocal coefficients. The most solid example of the theory
incorporating the mechanism (\ref{9.3}) is the
Dvali-Gabadaze-Porrati (DGP) model of brane-induced gravity
\cite{DGP} which is very interesting due to the fact that it
naturally contains the mechanism of the cosmological acceleration
\cite{BinDefEllLang}. However, this model turned out to be
suffering from the strong-coupling problem
\cite{A-HamGeSch,scale1,scale2} which invalidates the curvature
expansion theory, makes it inefficient without the fundamental
UV completion at the quantum level and features
the van Dam-Veltman-Zakharov (VDVZ) discontinuity problem
\cite{VDVZ}. Though this problem does not indicate the physical
inconsistency of the underlying theory and is likely to be
circumvented in tree-level applications by accounting for
nonlinear effects \cite{Vainshtein} or using the relativistic expansion
\cite{Kaloper}, in quantum domain it rises at full height,
because of the essentially perturbative nature of Feynman loop
expansion. Therefore, its status becomes very important for
possible generalizations of the Einstein theory.

The goal of this paper is to discuss the most general version of
the covariant nonlocal action for long-distance modifications of
Einstein theory, suggested in \cite{nonloc}. It turns out that in
the lowest (quadratic in curvature) order it is determined by two
nonlocal operators, kernels of Ricci and scalar curvature
quadratic forms, and also may contain additional term responsible
for the strong-coupling problem. The strongly coupled mode enters
this term linearly like the Lagrangian multiplier of the auxiliary
constraint on metric field and cannot be perturbatively integrated
out in terms of metric variables. After discussing this in Sect.2
we demonstrate generic structure of the action on the examples of
Fierz-Pauli and DGP models, consider the details of VDVZ and
strong-coupling problems and review their generalizations which allow
one to circumvent these problems. In particular, in Sect.5 we
review a recently suggested constrained perturbation theory in
the DGP model \cite{Gabadadze} and show that it turns out to be
a noncovariant modification of the original DGP model.
In the concluding section we briefly discuss the gauge independent
status of the VDVZ and strong-coupling problems and
outline the manifestly gauge-invariant technique for classical
and quantum braneworld models.

\section{Covariant nonlocal action for infrared modifications of
Einstein theory}

The idea of replacing the gravitational constant by a function of
the covariant d'Alember\-tian
$\Box=g^{\alpha\beta}\nabla_\alpha\nabla_\beta$, according to
\cite{A-HDDG} consists in the following modification of the left
hand side of Einstein equations
    \begin{eqnarray}
    \frac{M^2(\Box)}{16\pi}
    \left(R_{\mu\nu}-\frac12
    g_{\mu\nu}R\right)=\frac12\,T_{\mu\nu},  \label{9.8}
    \end{eqnarray}
where the nonlocal Planck mass is a function of the dimensionless
combination of $\Box$ and additional length scale $L$, interpolating
between the Planck constant for matter sources of small size $\ll L$
and long distance constant $G_{LD}=G(0)$,
    \begin{eqnarray}
    \frac1{G(\Box)}\equiv M^2(\Box)=M_P^2\,
    \big(1+{\cal F}(L^2\Box\,)\big).             \label{9.9}
    \end{eqnarray}
If the function of $z=L^2\Box$ satisfies the conditions
${\cal F}(z)\to 0$ at $z\gg 1$ and
${\cal F}(z)\to {\cal F}(0)\gg 1$ at $z\to 0$, then the infrared
modification is inessential for processes varying in spacetime
faster than  $1/L$ and vicy versa --- becomes significant ---
for slow phenomena with wavelengths $\sim L$ and higher.

An obvious difficulty with the construction of the above type is
that for any nontrivial operator ${\cal F}(L^2\Box\,)$ the left hand
side of (\ref{9.8}) does not satisfy the Bianchi identities and
cannot be obtained by varying the covariant action. In particular,
a naive attempt to modify the gravitational action according to
        \begin{eqnarray}
        M_P^2\,\int dx\,g^{1/2} R\,
    \Rightarrow\,
        \int dx\,g^{1/2} M^2(\Box) R=M^2(0)\,  \label{9.7}
        \int dx\,g^{1/2} R
        \end{eqnarray}
is meaningless, because after the integration by parts the
covariant d'Alembertian when acting to the left (on $1$)
picks up its zero mode and the nonlocal operator in all regimes
reduces to its infrared value $M_P^2(0)$.

This problem can be solved within the assumption of weak field
approximation implying that the equation (\ref{9.8}) represents
only the first linear term of the perturbation expansion in
powers of the curvature. Its left hand side must include higher
orders in the curvature, the nonlocal gravitational action
$S_{NL}[\,g_{\mu\nu}\,]$ generating the modified equations
according to
    \begin{eqnarray}
    \frac{\delta S_{NL}[\,g\,]}{\delta g_{\mu\nu}(x)}=
    \frac{M^2(\Box)}{16\pi}\,g^{1/2}\,
    \left(R^{\mu\nu}-\frac12
    g^{\mu\nu}R\right)+{\rm O}\,[\,R_{\mu\nu}^2\,]. \label{9.10}
    \end{eqnarray}

To obtain the leading term of $S_{NL}[\,g_{\mu\nu}\,]$, this
equation can be functionally integrated in the explicit form
\cite{nonloc} with the aid of the covariant curvature expansion
technique of \cite{CPT}. The essence of this technique consists in
the possibility to convert noncovariant series in powers of
gravitational perturbations $h_{\mu\nu}$ into the series of
spacetime curvature and its derivatives with the covariant
nonlocal coefficients. The starting point is the expansion of
the Ricci tensor
    \begin{eqnarray}
    &&R_{\mu\nu}=-\frac12\,\Box\,h_{\mu\nu}+\frac12\,\Big(
    \nabla_\mu F_\nu+
    \nabla_\nu F_\mu\Big)
    +{\rm O}\,[\,h_{\mu\nu}^2\,],         \label{9.12}
    \end{eqnarray}
($F_\mu\equiv \nabla^\lambda h_{\mu\lambda}-\frac12\,\nabla_\mu h$
is the linearized de Donder-Fock or harmonic gauge), which can be
solved by iterations with respect to $h_{\mu\nu}$ in the form of
the nonlocal expansion in the curvature, beginning with
    \begin{eqnarray}
    h_{\mu\nu}=-\frac2{\Box}R_{\mu\nu}
    +\nabla_\mu f_\nu+\nabla_\nu f_\mu
    +{\rm O}\,[\,R_{\mu\nu}^2\,].        \label{9.13}
    \end{eqnarray}
Here $\nabla_\mu f_\nu+\nabla_\nu f_\mu$ reflects the gauge
arbitrariness in this solution originating from the terms with
a harmonic gauge in (\ref{9.12}).

As a result, the nonlocal action generating (\ref{9.10}) begins
with the quadratic order in the curvature \cite{nonloc}
    \begin{equation}
    S_{NL}[\,g_{\mu\nu}]=
    -\frac1{16\pi}\,\int dx\,g^{1/2}
    \left\{\Big(R^{\mu\nu}
    -\frac12 g^{\mu\nu}R\Big)
    \frac{M^2(\Box)}{\Box}\,
    R_{\mu\nu}
    +{\rm O}[R_{\mu\nu}^3]\right\}.  \label{9.17}
    \end{equation}
Interestingly, in the simplest case of constant $M^2(\Box)=M_P^2$
it should reproduce the Einstein-Hilbert action, which looks
puzzling because it does not at all contain the term linear in
the curvature. The explanation of this paradox consists in the
observation that the Einstein action in the asymptotically-flat
spacetime with the metric behaving as
    $g_{\mu\nu}=\delta_{\mu\nu}+h_{\mu\nu}$,
    $h_{\mu\nu}={\rm O}\,\left(1/|x|^{2}\right)$, $|x|\to\infty$,
includes the Gibbons-Hawking surface integral over spacetime
infinity
    \begin{equation}
    S_E[\,g_{\mu\nu}]=-\frac{M_P^2}{16\pi}\,
    \int dx\,g^{1/2}\,R(\,g\,)+
    \frac{M_P^2}{16\pi}\,
    \int_{|x|\to\infty} d\sigma^\mu\,
    \big(\partial^\nu
    h_{\mu\nu}-\partial_\mu h\Big).    \label{9.11}
    \end{equation}
This integral can be transformed to the form of the volume
integral of $\partial^\mu(\partial^\nu
h_{\mu\nu}-\partial_\mu h)$ --- linear in $h_{\mu\nu}$ part of the
scalar curvature --- and, with the aid of (\ref{9.13}), covariantly
expanded in powers of the curvature. Up to quadratic terms inclusive
this expansion has the form \cite{nonloc}
    \begin{equation}
    \int_{|x|\to\infty}\!\! d\sigma^\mu
    \big(\partial^\nu
    h_{\mu\nu}-\partial_\mu h\Big)=
    \int dx\,g^{1/2}\left\{R
    -\Big(R^{\mu\nu}
    -\frac12\,g^{\mu\nu}R\Big)\frac1{\Box}R_{\mu\nu}
    +...\right\}.
    \end{equation}
Therefore, on substituting to (\ref{9.11}) linear in Ricci scalar
terms cancel out, and the quadratic terms reproduce the expression
(\ref{9.17}) with the coefficient $M^2(\Box)=M_P^2$ which can be
pulled out of the integral. The result is the {\it nonlocal} form
of the {\it local} Einstein action \cite{brane,nlbwa,nonloc},
    \begin{equation}
    S_E[\,g_{\mu\nu}]=
    -\frac1{16\pi}\,\int dx\,g^{1/2}
    \left\{\Big(R^{\mu\nu}
    -\frac12 g^{\mu\nu}R\Big)
    \frac{M_P^2}{\Box}\,
    R_{\mu\nu}
    +{\rm O}[R_{\mu\nu}^3]\right\}.  \label{9.17a}
    \end{equation}
The fact that this action begins with the quadratic order in
$R_{\mu\nu}\sim h_{\mu\nu}$ obviously corresponds to the theory of
massless spin-2 field. Less trivial is the nonlocality of this
action, which is the price one should pay for the manifest
covariance of this expansion in contrast to the local in terms of
$h_{\mu\nu}$, but noncovariant action for symmetric spin-2 tensor
field.

Thus the action (\ref{9.17}) serves as a direct realization of the
idea of nonlocal gravitational coupling constant --- replacement
of $M_P^2$ in the nonlocal version of the Einstein action
(\ref{9.17a}) by the operator (\ref{9.9}) as suggested in
\cite{nonloc}. However, there is a question (that was not
addressed in \cite{nonloc}) to what an extent this infrared
modification is generic even in the quadratic order approximation?
It is obvious that in the general case the operator kernels in the
quadratic Ricci tensor and scalar forms can be different, so that
the generalization of (\ref{9.17}) takes the form
    \begin{equation}
    S_{N\!L}[\,g_{\mu\nu}]=
    -\frac1{16\pi}\int dx\,g^{1/2}\left\{
    R^{\mu\nu}\frac{M_1^2(\Box)}{\Box}R_{\mu\nu}
    -\frac12\,R
    \frac{M_2^2(\Box)}{\Box}R
    +{\rm O}\,[\,R_{\mu\nu}^3\,]\right\},  \label{9.18}
    \end{equation}
where the two nonlocal "masses" should tend to one and the same
Planckian limit $M_P$ only in the short and intermediate
distance domain,
    \begin{equation}
    M_{1,2}^2(\Box)\to M_P^2,\,\,\,\, \Box\gg 1/L. \label{9.100}
    \end{equation}

The last requirement implies the recovery of the Einstein general
relativity at intermediate distances including, in particular, the
absence of the VDVZ discontinuity problem
--- correct tensor structure of the gravitational potential. For
the theory (\ref{9.18}) the linear potential generated by the
conserved matter sources $T_{\mu\nu}$ has the form
    \begin{eqnarray}
    h_{\mu\nu}=-\frac{16\pi}{M_1^2(\Box)\Box}\,
    \left(T_{\mu\nu}
    -\frac12\,\alpha(\Box)\,\eta_{\mu\nu}\,T\right)
    +\partial_\mu\xi_\nu+\partial_\nu\xi_\mu.     \label{9.16}
    \end{eqnarray}
Here the longitudinal terms correspond to gauge arbitrariness and
the operator coefficient $\alpha(\Box)$ equals
    \begin{eqnarray}
    \alpha(\Box)=\frac{2M_2^2(\Box)
    -M_1^2(\Box)}{3M_2^2(\Box)-2M_1^2(\Box)}.   \label{9.24}
    \end{eqnarray}
Generically, it should take the general relativistic value
$\alpha=1$ only in the limit $M_1^2(\Box)\to M_2^2(\Box)$
accounting for the restriction (\ref{9.100}).

Expression (\ref{9.18}) seems to give the most generic form of the
gravitational theory in weak field approximation, dictated by
general covariance. However, this is also not the end of the story
--- infrared limit of the theory together with spin-2
polarization can accommodate additional degrees of freedom that
are not taken into account in this expression. One would think
that, under a natural assumption that these degrees of freedom are
not directly coupled to matter, they can be integrated out, which
reduces to additional contributions to $M_1^2(\Box)$ and
$M_2^2(\Box)$. However, such a reduction is always possible except
the case when the equations of motion for additional fields cannot
be solved for them in terms of the metric. This happens when these
fields enter the action linearly and play the role of Lagrange
multipliers of certain combinations of metric variables.
Unfortunately, such fields persist at higher orders of
perturbation theory and, thus, give rise to the problem of the low
strong-coupling scale. This precludes from using the conventional
perturbation theory in a wide range of distances very well probed
by numerous Kavendish type and solar system experiments. Below we
demonstrate this phenomenon on the examples of the Fierz-Pauli
(FP) theory and the effective 4-dimensional theory of brane
induced gravity of Dvali-Gabadadze-Porrati (DGP).

\section{Fierz-Pauli model and VDVZ problem} The simplest
infrared modification of the Einstein theory is represented by the
Fierz-Pauli model of massive tensor field. It is described by the
quadratic part of the Einstein action (\ref{9.11}) modified by the
{\it noncovariant} mass term on the {\it flat}-spacetime
background
    \begin{eqnarray}
    &&S_{\rm mass}[\,g_{\mu\nu}]=-\frac{M_P^2}{16\pi}\,
    \int d^4x\,\left(\,\frac{m^2}4\,h_{\mu\nu}^2
    -\frac{m^2}4\,h^2\right),                 \label{9.19}\\
    &&h_{\mu\nu}\equiv g_{\mu\nu}
    -\eta_{\mu\nu},\,\,\,\,h\equiv
    \eta^{\mu\nu}h_{\mu\nu}.
    \end{eqnarray}
This is the only Lorentz-covariant combination of mass terms which
guarantees the absence of ghosts in the theory. Linear equations of
motion in this model have the form
    \begin{eqnarray}
    R_{\mu\nu}-\frac12\,\eta_{\mu\nu}\,R
    +\frac{m^2}2\,(h_{\mu\nu}-\eta_{\mu\nu}h)
    = 8\pi G_4\,T_{\mu\nu},                           \label{9.20}
    \end{eqnarray}
where under $R_{\mu\nu}$ we understand the part of Ricci tensor
linear in the graviton field. Differentiating this equation and
taking into account the linearized Bianchi identity and the
concservation of $T_{\mu\nu}$ we obtain the "gauge" for
$h_{\mu\nu}$
    \begin{eqnarray}
    \partial^\mu(h_{\mu\nu}-\eta_{\mu\nu}h)=0,  \label{9.21}
    \end{eqnarray}
which as a corollary has the vanishing of the linearized curvature
scalar
    \begin{eqnarray}
    R=\partial^\mu\partial^\nu h_{\mu\nu}-\Box h=0  \label{9.22}
    \end{eqnarray}
(note that this equation is satisfied even for nonvanishing trace of
matter stress tensor). As a result the gravitational field generated
by the matter source takes the form
    \begin{eqnarray}
    h_{\mu\nu}= -16\pi
    G_4\,\frac1{\Box-m^2}\,\left(T_{\mu\nu}
    -\frac13\eta_{\mu\nu}T\right)
    +\partial_\mu\xi_\nu+\partial_\nu\xi_\mu          \label{9.23}
    \end{eqnarray}
up to the longitudinal terms which do not couple to the conserved
matter sources. Since the FP theory is not gauge invariant, these
terms are not arbitrary and determined by the fixed vector
    \begin{eqnarray}
    \xi_\mu=-\frac{8\pi G_4}{3m^2}
    \frac1{\Box-m^2}\partial_\mu T,     \label{9.23a}
    \end{eqnarray}
which, in particular, guarantees the validity of Eq. (\ref{9.21}).

Tensor structure of massive graviton field here differs from the
general relativistic case --- the trace of $T_{\mu\nu}$ in
Einstein theory has a coefficient 1/2 rather than 1/3 in FP model.
This discrepancy remains also in the limit of vanishing mass and
constitutes the VDVZ discontinuity problem \cite{VDVZ}: massless
limit of FP model does not reproduce the predictions for massless
graviton of Einstein theory. This problem originates from the
additional degree of freedom which is missing in Einstein theory
and, on the contrary, exists in FP model for all values of the
graviton mass. From the viewpoint of the general framework for
infrared modifications of Einstein theory, this degree of freedom
should enter the action (\ref{9.18}) as a Lagrange multiplyer
responsible for the auxiliary equation (\ref{9.22}). Without it
the equations of motion following from (\ref{9.18}) do not
reproduce the tensor structure (\ref{9.23}) with any choice of
operators $M_1^2(\Box)$ and $M_2^2(\Box)$. Indeed, the coefficient
(\ref{9.24}) takes a numerical value of FP model,
$\alpha_{FP}=1/3$, only in a singular limit $M_1^2(\Box)\to 0$. On
the contrary, the inclusion of the constraint (\ref{9.22}) in
the action (\ref{9.18}) with a Lagrangian multiplier --- additional
scalar field --- improves the situation. It leads to the equations of
motion which yield as a solution the gravitational potential
(\ref{9.23}) for the following choice of nonlocal operators
    \begin{eqnarray}
    M_1^2(\Box)=M_2^2(\Box)=\frac1{G_4}\,
    \frac{\Box-m^2}{\Box}.                         \label{9.25}
    \end{eqnarray}


\section{DGP model and the strong coupling problem} Effective
long-distance modification of the Einstein theory is also provided
by the DGP model of brane induced gravity. Its fundamental action
includes the five-dimensional (bulk) Einstein term of the metric
$G_{AB}(X)$, $A=0,1,2,3,5$, and the Einstein term of the metric
$g_{\mu\nu}(x)$, $\mu=0,1,2,3$, on the 4-dimensional brane viewed
either as a boundary of the 5D bulk or as a fixed point of
$Z_2$-identification on the 5D orbifold
    \begin{eqnarray}
    &&S_{DGP}[\,G_{AB}(X)\,]=
    \frac1{16\pi G_5}\int d^5X\,
    G^{1/2}\,R^{(5)}(\,G_{AB}\,)
    \nonumber\\
    &&\qquad\qquad\qquad\qquad\qquad
    +\int d^4x\,g^{1/2}\left(\frac1{8\pi G_5}[K]
    +\frac1{16\pi G_4}\,R(\,g_{\mu\nu}\,)\right).     \label{10.25}
    \end{eqnarray}
The brane term is accompanied by the Gibbons-Hawking surface
integral of the trace of extrinsic curvature of the brane $[K]$
(analogous to the one at infinitely remote boundary written in
(\ref{9.11}) in the noncovariant form). The 4-dimensional
gravitational constant here $G_4=1/M_4^2$ is essentially different
from the 5-dimensional one $G_5=1/M_5^3$. They both determine two
different scales of the DGP model --- $M_4$ typically of the
Planckian value and the crossover scale to the infrared regime
    \begin{eqnarray}
    m=\frac{2M_5^3}{M_4^2}        \label{10.11}
    \end{eqnarray}
(usually identified with the Hubble scale of the Universe
$m=H=10^{-28}$ cm$^{-1}$).

As is known, the DGP model is analogous to the FP theory with the
mass term more soft in the infrared, $m^2\Rightarrow
m\sqrt{-\Box}$, in which the role of $m$ is played by the scale
(\ref{10.11}). To show this, one can build the effective brane
action which follows from (\ref{10.25}) by integrating out the
bulk metric $G_{AB}(X)$ subject to fixed boundary data on the
brane. This can be done perturbatively. For this purpose expand
the fundamental action in gravitational perturbations
    \begin{eqnarray}
    G_{AB}(X)=\eta_{AB}+H_{AB}(X),               \label{9.26}
    \end{eqnarray}
solve the linear equations for $H_{AB}(X)$ in the bulk and
substitute the result back into the quadratic part of the action.
We shall work in the coordinate system in which the brane is located
at the fixed value of the fifth coordinate, $X^5\equiv y=0$. For
gauge fixing in the bulk add to the action (\ref{10.25}) the
gauge breaking term quadratic in the linearized de Donder-Fock
gauge
    \begin{eqnarray}
    &&S_{\rm gauge}[\,H_{AB}]
    =-\frac{M^3_5}{16\pi}\,
    \frac12\int d^5X\,\eta^{AB}F_A F_B,  \label{9.27}\\
    &&F_A=\partial^B\!H_{AB}
    -\frac12\,\partial_A H.               \label{9.28}
    \end{eqnarray}
In this gauge the bulk equations of motion are most simple and
form the following boundary value problem
    \begin{eqnarray}
    &&\Box_5 H_{AB}(X)=0,                        \label{9.29a}\\
    &&H_{AB}(x,y)\,\big|_{y=0}=h_{AB}(x),\,\,\,
    h_{AB}(x)\equiv
    \big(h_{\mu\nu}(x),N_\mu(x),h_{55}(x)\big).    \label{9.29}
    \end{eqnarray}
On a flat background $\Box_5=\Box+\partial_y^2$, and the solution
of this problem nonsingular at the bulk infinity can be written
down in the following elegant form \cite{scale1}
    \begin{eqnarray}
    H_{AB}(x,y)=e^{-y\Delta}\,h_{AB}(x)   \label{9.30}
    \end{eqnarray}
in terms of the auxiliary operator
    \begin{eqnarray}
    \Delta=\sqrt{-\Box}         \label{9.31}
    \end{eqnarray}
(the case of Lorentzian spacetime we treat as the analytic
continuation from the Euclidean space where the operator $\Box$
is negative definite, so that $H_{AB}(x,y)$ vanishes at $y\to\infty$).

Substituting the obtained solution in the 5D part of the DGP
action (\ref{10.25}) $S_5[\,G_{AB}]$ (with the five-dimensional
curvature and the Gibbons-Hawking integral) and taking into account
the gauge breaking term (\ref{9.27}) one gets \cite{scale1}
    \begin{eqnarray}
    &&S_5[\,G_{AB}]+ S_{\rm gauge}[\,H_{AB}]=
    \frac{M^2_4}{16\pi}\,\frac{m}4\,
    \int d^4x\,\left(-{\tilde h}^{\mu\nu}
    \Delta{\tilde h}_{\mu\nu}
    +\frac12{\tilde h}\Delta{\tilde h}\right.  \nonumber\\
    &&\qquad\qquad\qquad\qquad\qquad\qquad\qquad\qquad
    +\left.{\tilde h}\Delta h_{55}
    -\frac12 h_{55}\Delta h_{55}\right),    \label{9.32}
    \end{eqnarray}
where $m=2M_5^3/M_4^2$ is the DGP scale (\ref{10.11}) and
${\tilde h}_{\mu\nu}$ is the following combination of the metric
induced on the brane and shift functions in the fifth dimension
$G_{5\mu}=N_\mu$
    \begin{eqnarray}
    {\tilde h}_{\mu\nu}=h_{\mu\nu}
    +\frac1\Delta\,(\partial_\mu N_\nu
    +\partial_\nu N_\mu).                     \label{9.33}
    \end{eqnarray}
Note that this combination is gauge invariant with respect to
the four-dimensional coordinate transformations
    \begin{eqnarray}
    \delta_\xi h_{\mu\nu}=\partial_\mu \xi_\nu
    +\partial_\nu \xi_\mu,\,\,\,
    \delta_\xi N_\mu=-\Delta \xi_\mu,\,\,\,
    \delta_\xi h_{55}=0.              \label{9.34}
    \end{eqnarray}
In their turn, these transformations represent the restriction to
the brane of the residual gauge transformations in the bulk
$\delta_\Xi H_{AB}=\partial_A\Xi_B +\partial_B \Xi_A$,
    \begin{eqnarray}
    \Xi^\mu(x,y)=e^{-y\Delta}\,\xi^\mu(x),\,\,\,\,
    \Xi^5(x,y)=0,                               \label{9.34a}
    \end{eqnarray}
which, because of $\Box_5\Xi^A(x,y)=0$, leave invariant the de
Donder gauge (\ref{9.28}) and do not shift the brane from its
location at $y=0$, .

Thus, as expected, the effective action (\ref{9.32}) induced from
the bulk is invariant with respect to four-dimensional
transformations, but this invariance is actually realized by means
of the Stueckelberg fields $N_\mu$, usually introduced by hands
for the covariantization of the noninvariant action. By varying
(\ref{9.32}) with respect to $N_\mu$, these fields can be excluded
in terms of the metric variables
    \begin{equation}
    N_\mu=\frac1\Delta\left(\partial^\nu h_{\mu\nu}
    -\frac12\partial_\mu h
    -\frac12\partial_\mu h_{55}\right),
    \end{equation}
which leads to the manifestly invariant expression for
${\tilde h}_{\mu\nu}$ in terms of the linearized Ricci tensor
    \begin{eqnarray}
    {\tilde h}_{\mu\nu}=
    -2\frac1\Box R_{\mu\nu}
    +\frac{\partial_\mu
    \partial_\nu}\Box h_{55}. \label{9.35}
    \end{eqnarray}
Their substitution to the bulk action leads to
    \begin{equation}
    S_5+ S_{\rm gauge}=
    \frac{M^2_4}{16\pi}\,m\,
    \int d^4x\,\left(-R^{\mu\nu}
    \frac\Delta{\Box^2}R_{\mu\nu}
    +\frac12R\frac\Delta{\Box^2}R
    -R\frac\Delta{\Box}h_{55}
    \right),                        \label{9.36}
    \end{equation}
where the scalar variable $h_{55}$ can no longer be excluded in terms
of the metric field --- the situation discussed above.

Adding (\ref{9.36}) to the four-dimensional part of the DGP action,
rewritten in the nonlocal form (\ref{9.17}), we finally obtain
the quadratic part of the effective action on the brane
    \begin{equation}
    S_{DGP}^{\rm eff}[\,g_{\mu\nu},\Pi\,]
    =-\frac{M_4^2}{16\pi}\int dx\,g^{1/2}
    \left\{\Big(R^{\mu\nu}
    -\frac12 g^{\mu\nu}R\Big)
    \frac{\Box-m\Delta}{\Box^2}\,
    R_{\mu\nu}+m\Pi R\right\},  \label{9.37}
    \end{equation}
where the Lagrangian multiplier to the scalar curvature $\Pi$ is
related to $h_{55}$ component of the 5D metric according to
    \begin{eqnarray}
    h_{55}=-2\Delta\Pi. \label{9.38}
    \end{eqnarray}
The variable $\Pi$ was introduced in \cite{scale1} as a
longitudinal part of the 5D shift function
$N_\mu=\partial_\mu\Pi+N_\mu'$. It parameterizes the brane bending
in the form of the 5D diffeomorphism of the bulk metric
    \begin{eqnarray}
    &&\delta_\Xi H_{5\mu}
	=\partial_\mu\Xi_5, \label{9.101a}\\
    &&\delta_\Xi H_{55}=2\,
	\partial_y\Xi_5=-2\Delta\Pi,  \label{9.101}
    \end{eqnarray}
with the vector field $\Xi^A(x,y)=\delta^A_5\,e^{-y\Delta}\Pi(x)$.
Similarly to (\ref{9.34a}) this diffeomorphism does not break the
de Donder gauge in the bulk, but it shifts the brane in the
direction of the fifth coordinate by $\Xi^5(x,0)=\Pi(x)$ and,
therefore, is not the symmetry of the action. It does not manifest
itself in the bulk, and its effect reduces to the contribution on
the brane, which begins with the local term $m\Pi R$.

Thus, the 4-dimensional effective action of the DGP model does not
take the form (\ref{9.18}), because it contains auxiliary
constraint with the Lagrange multiplier which cannot be expressed
in terms of metric variables. Therefore, in the linear
approximation the DGP model on the brane is effectively described
by the Fierz-Pauli theory with the nonlocal mass term of the form
(\ref{9.36}), generated from the bulk. In essence, the expression
(\ref{9.36}) turns out to be the covariant completion of this term
(\ref{9.19}) with the infrared soft mass \footnote{Covariant
structures of such a type as a realization of the nonlocal
cosmological "constant" were also discussed in context of the
renormalization theory in \cite{nneag,ShapGorb}.}
$\sqrt{m\Delta}$. Linearized gravitational potential of matter
source in this model is analogous to (\ref{9.16})
    \begin{eqnarray}
    h_{\mu\nu}= -16\pi
    G_4\,\frac1{\Box-m\Delta}\,\left(T_{\mu\nu}
    -\frac13\eta_{\mu\nu}T\right)
    +\partial_\mu\xi_\nu+\partial_\nu\xi_\mu.          \label{9.38a}
    \end{eqnarray}
Interaction of matter on the brane is determined here by the
propagator which for small spacetime intervals $|x|\ll L=1/m$,
$|\Box|\gg m^2$, obviously coincides with the 4-dimensional one.
On the contrary, beyond the crossover scale (\ref{10.11}) the
coupling becomes five-dimensional. This is usually interpreted as
a gravitational leakage to the bulk --- 4D graviton is metastable
and decays within a lifetime $L=1/m$ similarly to the
Gregory-Rubakov-Sibiryakov model \cite{GRS}. In contrast to GRS
model which suffers from ghost negative-energy states
\cite{PilRatZaf,DubRub,DubLib}, the DGP model (like FP theory) is
ghost-free.

Similarly to FP model, the constraint term of (\ref{9.37}) results
in the VDVZ problem --- tensor structure of the gravitational
potential (\ref{9.38a}) differs from general relativity and
corresponds to to Fierz-Pauli theory in all ranges of distances.
Note that kinematically the variable $\Pi$ of brane embedding into
the bulk is analogous here to the radion mode in the
Randall-Sundrum model \cite{GarrigaTanaka}, which guarantees there
the recovery of the general relativistic structure. Here, however,
this mode does not cope with this task and the DGP model suffers
from the VDVZ problem \cite{DGP}.

Another consequence of the constraint term in the DGP action
(\ref{9.37}) is the problem of low strong-coupling scale. Point is
that the variable $\Pi$ has a nature of the Lagrange multiplier
only in the quadratic order, while in higher order terms of the
weak field expansion it enters the action nonlinearly and gives
rise to composite operators built of the powers of $\Pi$, $N_\mu$,
$h_{\mu\nu}$ and their derivatives. On the other hand, its kinetic
term in the quadratic order originates entirely from mixing with
$h_{\mu\nu}$ (in view of its Lagrangian multiplier nature in
$M_P^2\,m\,\Pi R\sim M_P^2\,m\,\partial\Pi\,\partial h$) and is
small because of smallness of $m$. As a result, after the
diagonalization of the full quadratic term the $\Pi$-mode acquires
the kinetic term $\sim M_P^2 m^2 (\partial\Pi)^2$, and the
transition to the canonically normalized field $\hat\Pi$,
$\Pi=\hat\Pi/(m M_P)$, when expanding in $\Pi$ gives rise to
higher and higher negative powers of small quantity $m$
\cite{A-HamGeSch}. Then, the composite operators of high
dimensionalities become suppressed by the factors of the form
$1/M_P^p\,m^q$ and get strong at the scale
$\Lambda_{p,\,q}=(M_P^p\,m^q)^{1/(p+q)}$. As shown in
\cite{scale1,NicRat}, the lowest scale occurs for the cubic in
$\Pi$ interaction
    \begin{eqnarray}
    -\frac1{\Lambda^3}\int d^4x\,(\partial\Pi)^2\Box\Pi,
    \end{eqnarray}
and its value
    \begin{equation}
    \Lambda=
    (m^2 M_P)^{1/3}\sim (1000 \,\mbox{km})^{-1}  \label{9.41}
    \end{equation}
is much lower than the submillimeter scale $(0.02 \mbox{mm})^{-1}$
up to which the Newton law is verified by high precision table-top
experiments \cite{Gundlach}. Analogous situation is well known for
the nonlinear FP model (with the full Einstein term and quadratic
mass term) --- its strong-coupling scale being equal to
$\Lambda_5=(m^4 M_P)^{1/5}$ \cite{A-HamGeSch}\footnote{At the
classical level the strong-coupling and VDVZ problems in the FP
model may apparently be circumvented by infinite resummation of
nonlinear terms \cite{Vainshtein}. However, at the quantum level
the theory has the strong coupling scale \cite{Aubert} which at
best can be raised to $(m^2 M_P)^{1/3}$ by the inclusion of
higher-dimensional operators \cite{A-HamGeSch}.}.

The recovery of Einstein phase in the DGP theory (with the correct
tensor structure of the propagator) can be attained by its
synthesis with the Randall-Sundrum model. If the brane has a
positive tension fine tuned to the negative cosmological constant
in the bulk, then in the perturbative domain below the strong
coupling scale, $\Box\ll\Lambda^2$, the gravitational potential
(\ref{9.16}) has operator functions \cite{TanakaDGP}
    \begin{eqnarray}
    &&M_1^2(\Box)=M_4^2\left[\,1+
    \frac{m\,K_1(l\Delta)}{\Delta\,K_2(l\Delta)}\,\right],\\
    &&\alpha(\Box)=\frac2{2+m\,l}+\frac23\,\frac{lm}{2+l\,m}\,
    \left[\,1+\frac{K_1(l\Delta)}{l\Delta\,K_2(l\Delta)}\,\right],
    \end{eqnarray}
in terms of McDonald functions of the first and second order,
$K_{1,2}(x)$, and curvature radius $l$ of the AdS bulk. In the
distance range $1/m\gg1/\Delta\gg l$ this potential describes the
4-dimensional general relativistic law with $\alpha(\Box)\simeq 1$
and effective Planck mass $M_P^2=M_4^2\,(1+lm/2)\simeq M_4^2$
\cite{TanakaDGP}. Thus, within the hierarchy of the horizon (DGP)
and the AdS scales $1/m=1/H\gg l$ this model does not suffer from
the VDVZ problem. Actually, this is a generalization of the
well-known result that this problem is absent for spin-2 massive
field in (A)dS spacetime \cite{Higuchi,Porrati-disc} with the
cosmological constant $\Lambda$ in the limit $m^2/\Lambda\to 0$.

Another generalization that helps to resolve the VDVZ problem
consists in increasing the number of extra dimensions in the DGP
model, which has a good motivation from string theory
\cite{bigNstring}. Correct tensor structure in this case is,
however, achieved by the price of ghost states of tachyon nature
\cite{DubRub} which indicates classical and quantum instabilities.
Two approaches were suggested to avoid these states, based on the
necessity to make the UV regularization for branes with
codimension $N>1$ \cite{reg}. The regula\-rization smearing the
brane was used in \cite{KolPorRomb} (together with the
$D=4+N$-dimensional scalar curvature replacing on the brane the
4-dimensional one) and the model was shown to be ghost and tachyon
free, and, moreover, its 5D version acquiring a big
strong-coupling scale $\Lambda_9=(M_5^7 m^2)^{1/9}\gg \Lambda_3$
much exceeding (\ref{9.41}) \cite{PorRomb}.

Another approach to DGP models with $N\geq 2$ \cite{soft} was
based on the interpretation of Green's functions poles different
from \cite{DubRub}. The unitarity of the theory was recovered by
noting that the ghost tachyon of \cite{DubRub} belongs to the
unphysical sheet of complex Mandelstam variable and, apparently,
can be removed by the sort of the Lee-Wick prescription in local
field theory with ghost states \cite{LeeWick}). The accompanying
loss of analyticity of the propagator results in the loss of
causality at the horizon scale $L=M_4/M_D^2$, which as advocated
in \cite{soft} does not contradict causality at small and
interme\-diate distance scales.

\section{Constrained DGP model}
VDVZ and strong coupling problems in the DGP model invalidate
the weak field expansion every\-where
except the ultrainfrared domain of energies $E$ below the horizon
scale $E\ll m=1/H_0$. This makes the flat-space perturbation
theory (and its covariantization according to the lines of Sect.2)
inefficient. Moreover, it turns out to be incomplete without the the
UV completion at the quantum level, because the infinite tower of
higher dimensional operators limits its predictive power even for
larger distances. Recently there were several attempts to save the
situation \cite{NicRat,Gabadadze}. One of them \cite{NicRat} is
based on the transition to the dynamically more safe non-flat
background on which the nonvanishing brane extrinsic curvature
$K_{\mu\nu}$ introduces big kinetic term for the brane bending
mode and, therefore, acts like a dilaton field controlling the
strength of its interaction at each spacetime point. More radical
is the suggestion of \cite{Gabadadze} to interpret the strong
coupling problem as a gauge artifact that can be
circumvented by a special modification of perturbation theory called
in \cite{Gabadadze} the constrained perturbation theory.

This modification consists in adding to the action (\ref{10.25}) the
full gauge-fixing term
    \begin{eqnarray}
    S_{\rm gauge}[H_{AB}]=-\frac{M_4^2}{16\pi}
	\int d^4x\,\frac{F_\mu^2}{2\sigma}
	+\frac{M_5^3}{16\pi}
	\int d^5X\left(\frac1{2\gamma}B_5^2
	-\frac1{2\beta}B_\mu^2\right),       \label{5.1}
    \end{eqnarray}
which includes the set of special 5-dimensional gauges in the bulk
    \begin{equation}
    B_\mu=\partial_\mu H_{55}
	-\partial^\alpha H_{\alpha\mu},
    \,\,\,B_5=\partial^\mu H_{\mu 5},     \label{5.2}
    \end{equation}
along with the 4-dimensional harmonic gauge for the intrinsic metric
$h_{\mu\nu}$ on the brane
    \begin{equation}
    F_\mu=\partial^\nu h_{\mu\nu}
	-\frac12\partial_\mu h.        \label{5.3}
    \end{equation}
From the viewpoint of gauge-fixing procedure the brane gauge term
is redundant, because the bulk term in (\ref{5.1}) alone
completely fixes all coordinate invariances of the theory\footnote{The
residual gauge transformations for the bulk gauge (\ref{5.2}),
$\delta H_{AB}=2\partial_{(A}\Xi_{B)}$, are locally parametrized by {\em one}
scalar field $\Xi_5$ propagating in the bulk, $\Box_5\Xi_5=0$,
$\Xi_\mu=\partial_\mu(1/\Box)\partial_5\Xi_5$, but the corresponding
diffeomorphism should not move the brane, $\Xi_5 |=0$. Therefore, with
this condition on the brane and zero boundary conditions at infinity
$\Xi_5(X)$ vanishes identicallty everywhere in the bulk and thus
does not leave any gauge freedom.}. Therefore
this term was interpreted in \cite{Gabadadze} as an additional
constraint extending beyond conventional gauge-fixing procedure. The
calculations show that for particular limiting values of "gauge-fixing"
parameters $\beta\to 0$, $\gamma\to 0$ and $\sigma=1$ the gravitational
potential of matter sources on the brane does not suffer from
strong coupling and VDVZ problems. In particular, in Eq.(\ref{9.23})
it has the operator coefficient $\alpha(\Box)$ interpolating
between the FP value $\alpha_{FP}=2/3$ in the deep IR regime
$\Delta\equiv\sqrt{-\Box}\ll m$ and the Einstein value $\alpha=1$
for short distances $\Delta\gg m$ and does not have terms singular
for $m\to 0$.

This result, however, can hardly be interpreted as a proper solution of
these problems in the DGP model. Rather than being the
constrained perturbation theory for this model, the suggestion of
\cite{Gabadadze} represents the modification of the model itself,
and this modification is not unique. Indeed, the theory with
the action (\ref{10.25}) modified by the term (\ref{5.1}) in the limit
    \begin{equation}
	\beta\to 0,\,\,\,\,\gamma\to 0,    \label{5.4}
    \end{equation}
but with an arbitrary parameter $\sigma$, has the following
gravitational potential on the brane,
$h_{AB}=H_{AB}|$, (in what follows the vertical bar denotes
the restriction of a bulk quantity to the brane)
   	\begin{eqnarray}
	&&h_{\mu\nu}=
	-\frac{16\pi}{M_4^2}\,\frac1{\Box-m\Delta}\left(\,T_{\mu\nu}-
	\frac12\,\alpha_\sigma(\Box)
	\eta_{\mu\nu} T\right)
	+\frac{16\pi}{M_4^2}\,\partial_\mu\partial_\nu
	\frac{\omega_\sigma(\Box)}\Box\,T,        \label{33}\\
	&&h_{55}=
	\frac{16\pi}{M_4^2}\,\frac12\frac{\Box-2\sigma m\Delta}
	{\Box-4m\Delta-3\sigma m^2}\,
	\frac1\Box\,T,                            \label{33a}\\
	&&h_{5\mu}=0                              \label{33b}
	\end{eqnarray}
with the operator coefficients (see the derivation in Appendix)
    	\begin{eqnarray}
    	&&\alpha_\sigma(\Box)=
	\frac{\Box-3m\Delta-2\sigma m^2}
	{\Box-4m\Delta-3\sigma m^2},   \label{9.104}\\
	&&\omega_\sigma(\Box)=
	\frac1{\Box-m\Delta}\,
	\frac{(1-\sigma)\,m\Delta}
	{\Box-4m\Delta-3\sigma m^2}.        \label{35}
    	\end{eqnarray}
The data (\ref{33})-(\ref{33b}) is propagated from the brane into the
bulk by the massless Klein-Gordon equation\footnote{This is a
corollary of the limit (\ref{5.4}). For nonvanishing $\beta$ and
$\gamma$ only the transverse-traceless, $H_{\mu\nu}^{TT}$, and conformal,
$\eta_{\mu\nu}\Phi$, sectors of $H_{AB}$ are subject to this propagation
law (see Eqs.(\ref{A.19})-(\ref{A.21}) in Appendix).} (\ref{9.29a})
according to (\ref{9.30}) and does not have dangerous parts blowing
at $m\to 0$.

The expression (\ref{33}) for on-brane
metric can be obtained from the effective action
of the form (\ref{9.18}) with the nonlocal form factors
    \begin{equation}
    M_1^2(\Box)=\frac1{G_4}\,
    \frac{\Box-m\Delta}{\Box},\,\,\,
    M_2^2(\Box)=\frac1{G_4}\,
    \frac{\Box-2m\Delta-\sigma m^2}{\Box}.       \label{9.103}
    \end{equation}
This action, in contrast to (\ref{9.37}), does not contain any
constraint term, because no strong coupling mode is present and
all variables express in terms of the 4-dimensional metric.
Since the constrained theory is not gauge-invariant, the longitudinal
term of the solution (\ref{33}) is not ambiguous just like in
the Fierz-Pauli model with (\ref{9.23a}). Therefore, this solution
can be regarded as obtained from the {\em covariant} effective
action (\ref{9.18}) in a particular $\sigma$-dependent gauge
	\begin{equation}
	\partial^\nu h_{\mu\nu}
	-\frac12\frac{\Box-2\sigma m\Delta}{\Box-\sigma m\Delta}
	\partial_\mu h=0.
	\end{equation}
This gauge interpolates between the Fierz-Pauli gauge (\ref{9.21})
in the infrared and the harmonic gauge for high energies, just like
the operator coefficient (\ref{9.104}) interpolates between the
corresponding Fierz-Pauli and Einstein values.

Thus, for generic nonsingular $\sigma$ the brane metric
(\ref{33}) has the Einstein high-energy limit and the (soft-mass)
Fierz-Pauli behavior at low energies. But at intermediate energies it
corresponds to the one-parameter family of physically inequivalent
theories. In particular, they are not equivalent to the original DGP model,
because their equations are spoiled by gauge terms which are nonvanishing
even on shell. For example, one of the bulk Einstein equations in the
DGP model implies the vanishing 4-dimensional scalar curvature,
${\cal E}_{55}^{(5)}\sim R^{(4)}=0$, while for (\ref{33}) it reads
    \begin{equation}
    R^{(4)}=-\frac{8\pi}{M_4^2}\,
	\frac1{\Box-4m\Delta-3\sigma m^2}\,T	    \label{5.5}
    \end{equation}
and goes to zero only in the singular limit $\sigma\to\infty$ (this
limit obviously corresponds to the original DGP model and is
marred by the strong-coupling and VDVZ problems).

The constrained theories depend not only on the $\sigma$, but
also on the bulk "gauge-fixing" parameters $\beta$ and $\gamma$.
This gauge dependence is an artifact
of introducing the redundant brane "gauge" in (\ref{5.1}).
As shown in Appendix, the Israel junction condition which
underlies the derivation of (\ref{33})-(\ref{33b}) establishes the
relation (\ref{A.33}) between this gauge $F_\mu$ and the
boundary value of the
bulk gauge $B_5\,|$. Only their combination (\ref{A.33}) is vanishing
on shell, so that separately $F_\mu(x)$ and $B_A(X)$ are
nonzero\footnote{Brane gauge $F_\mu$ can be separately put to zero
by demanding the continuity of $H_{5\mu}$-coefficients across
the brane\cite{LibRub}, but this requirement seems contrived, because
it does not follow from the geometrically invariant formulation of the
$Z_2$-symmetry of brane orbifolds.}. Therefore these constraints go
beyond a consistent gauge-fixing procedure and modify the original
model.

\section{Discussion and conclusions}
The above discussion shows that the strong-coupling and VDVZ problems
are very robust within covariant infrared modifications
of Einstein theory. In particular, for DGP model the VDVZ problem
is a direct corollary of the 5-dimensional Einstein equations.
The $(55)$-component of these equations implies the vanishing of the
4-dimensional scalar curvature (\ref{9.22}) which means that there
is no conformal term $\eta_{\mu\nu}\varphi$ in the transverse-traceless
decomposition of the metric potential on the brane
$h_{\mu\nu}= h_{\mu\nu}^{TT}+\partial_\mu\xi_\nu+\partial_\nu\xi_\mu$.
Therefore, up to gauge transformation terms the linearized
gravitational field is given by the transverse-traceless part of the
matter stress tensor, $h_{\mu\nu}\sim T_{\mu\nu}^{TT}$, and
this unambiguously fixes the $\alpha$-coefficient of (\ref{9.16}) at
the Fierz-Pauli value $2/3$, because
	\begin{equation}
	T_{\mu\nu}^{TT}=
	T_{\mu\nu}-\frac13\eta_{\mu\nu}T
	+\frac13\partial_\mu\partial_\nu\frac1\Box\,T.
	\end{equation}
Thus, the VDVZ problem in DGP model cannot be circumvented without
spoiling the basic dynamical equations in the bulk. This is
exactly what happens in the constrained DGP model of \cite{Gabadadze}.

The strong coupling problem also has a gauge-invariant origin, even
though the derivation of Sect.4 seems to point at the particular
harmonic gauge as a source of this difficulty.
Note that in the derivation of (\ref{9.37}) we could not exclude
via the variational equation the $h_{55}$ metric component which turned
out to be the strongly coupled mode (\ref{9.38}) --- the Lagrangian
multiplyer of the scalar curvature constraint in (\ref{9.37}). The
impossibility to solve the linearized Einstein equations in the bulk
for lapse and shift functions $N_A=G_{5 A}$ follows from the
degeneracy of the operator
    \begin{equation}
    S^{AB}\delta(X,Y)=\frac{\delta^2 S_{5}[\,G\,]}
	{\delta N_A(X)\,\delta N_B(Y)}.	    \label{6.1}
    \end{equation}
On flat space background it is really degenerate because its $(5A)$-column
is identically zero and the $(\mu\nu)$-block,
$S^{\mu\nu}=\partial^\mu\partial^\nu-\eta^{\mu\nu}\Box$, has as
a zero mode an arbitrary longitudinal vector $N_\nu=\partial_\nu\Pi$
--- the $(5\nu)$-component of the strong coupling mode (\ref{9.101a}).

At this point it is useful to compare the gauge (in)dependence
properties of the original DGP model and its constrained version
of Sect.5. The de Donder gauge $F_A$, (\ref{9.28}), used in Sect.4 is
qualitatively different from the gauge $B_A$ in (\ref{5.2}). It
belongs to the class of relativistic gauges for which the residual
transformations are locally parametrized by the full set of five
propagating modes, $\Box_5\Xi_A=0$, and the Faddeev-Popov operator
$\Box_5$ is dynamical for all components of $\Xi_A$. The fifth
component $\Xi_5$ is ruled out by Dirichlet boundary conditions
on the brane $\Xi_5|=0$ (because the movement of the brane is not
a gauge transformation), and the residual transformations reduce to
4-dimensional diffeomorphisms (\ref{9.34a}). Their fixation requires
four extra gauges imposed on the brane, and this is a main point
of departure from the special gauge (\ref{5.2}) of \cite{Gabadadze}.
In contrast to \cite{Gabadadze} these extra gauges
do not violate Ward identities and do not change the
physical contents of the model. This property is based on the fact
that, though the effective brane action (\ref{9.32}) is gauge dependent,
its gauge dependence gets eradicated by the transition to the
effective action (\ref{9.37}) with on-shell values of the functions
$N_A$. This is different from the constrained version of the DGP
model for which the effective form factors (\ref{9.103}) explicitly
depend on all -- bulk and brane -- "gauge-fixing" terms.

To clarify the properties of the action (\ref{9.32}) versus (\ref{9.37}),
note that the former was obtained as a fundamental
DGP action
    \begin{eqnarray}
    &&S_{DGP}^{\rm eff}[\,g_{\mu\nu},\,N_A\,]=
    S_{DGP}[\,G_{AB}\,]\,
    \Big|_{G_{AB}=G_{AB}[\,g_{\mu\nu},\,N_A\,]}  \label{9.105}\\
    &&g_{\mu\nu}=G_{\mu\nu}|\,,\,\,
    N_A=G_{5 A}|\,,
    \end{eqnarray}
evaluated on the solution of the bulk Einstein equations
$G_{AB}[\,g_{\mu\nu},N_A\,]$ subject to independent boundary
conditions for {\it all} five-dimensional metric coefficients on
the brane. The action functional for such boundary conditions
depends on the choice of gauge, because its on-shell
restriction is not complete regarding the boundary
values $N_A(x)$ of $G_{5A}(X)$-coefficients. Only their exclusion by
virtue of variational equations
    \begin{equation}
    \frac{\delta S_{DGP}^{\rm eff}}{\delta N_A(x)}=0   \label{9.106}
    \end{equation}
in terms of the metric $g_{\mu\nu}$, $N_A=N_A[\,g_{\mu\nu}]$,
makes the action gauge independent\footnote{In the relativistic
gauge $F_A(X)$, (\ref{9.28}), the
variational equations (\ref{9.106}) represent the requirement of
zero boundary data for these gauge
conditions, $F_A|=0$. In the bulk they satisfy the homogeneous
equation $\Box_5 F_A(X)=0$ (with the Faddeev-Popov operator
$\Box_5$). Therefore $F_A(X)=0$ everywhere in the bulk and the
on-shell contribution of the gauge-breaking term (\ref{9.27})
vanishes and makes the whole on-shell action gauge independent.}.
However, the strong-coupling problem prevents
from solving these equations for all
$N_A=N_A[\,g_{\mu\nu}]$ -- only its transverse $\mu$-components
can be excluded, and at least one scalar mode $\Pi$ remains
off shell among the arguments of (\ref{9.37}). Nevertheless,
the result turns out to be covariant and bulk-gauge independent,
though suffering from the strong coupling and VDVZ problems.

In essence, the gauge difficulties with these problems follow from
the lack of manifestly covariant and gauge independent formalism.
The on-shell reduction which makes the brane action
gauge independent does not resolve this problem. In particular,
the construction of brane-to-brane propagator requires the
off-shell exten\-sion of the brane action and, therefore, is sensitive
to the choice of gauge. Moreover, the off-shell
extension of the action is necessary at the quantum level,
where it should be integrated over the fields on the brane to
generate the full set of Feynman diagrams. This integral in which
the bulk integration is done first, while the integration over
brane fields is reserved for the last, schematically looks like
    \begin{eqnarray}
    &&\int
    D_5G_{AB}\,e^{-S_{DGP}[G_{AB}]}\,\big(...\big)\nonumber\\
    &&\qquad\qquad\qquad=
    \int D_4 g_{\mu\nu}\,D_4 N_A\,
    e^{-S_{DGP}^{\rm eff}[g_{\mu\nu},\,N_A]
    -\SGamma_{\rm loop}[g_{\mu\nu},\,N_A]}
    \,\big(...\big).                               \label{9.107}
    \end{eqnarray}
Here ellipses denote all the details of gauge-fixing procedure
(including the contribution of relevant Faddeev-Popov
determinants), the subscript of $D$ in the integration measure
indicates the spacetime dimensionality of fields
--- path integral over the 5-dimensional metric $G_{AB}(X)$ in
the bulk versus  4-dimensional fields $g_{\mu\nu}(x)$ and $N_A(x)$
on the brane. $\SGamma_{\rm loop}[\,g_{\mu\nu},\,N_A\,]$ denotes
the loop part of the brane effective action following from the
functional integration over bulk fields bounded by brane fields
$g_{\mu\nu}(x)$ and $N_A(x)$.

In this form the functional integral is not manifestly covariant,
because the integrand on the right-hand side is off shell and,
therefore, gauge dependent. Its overall gauge independence is not
manifest because it is attained only as a result of brane field
integration. However, the strategy of its calculation can be
improved by including the integration over $D_4 N_A$ into the
bulk, so that (\ref{9.107}) takes the form
    \begin{eqnarray}
    \int D_4 g_{\mu\nu}\,
    e^{-S^{\rm eff}[g_{\mu\nu}]
    -\SGamma_{\rm loop}[g_{\mu\nu}]}
    \,\big(...\big)                              \label{9.108}
    \end{eqnarray}
with the brane tree-level $S^{\rm eff}[\,g_{\mu\nu}\,]$ and loop effective
$\SGamma_{\rm loop}[g_{\mu\nu}]$ actions {\it gauge independent
off shell}. Such impro\-vement, the corresponding Feynman
diagrammatic technique and its gravitational Ward identities will
be presented in the coming paper \cite{QEAB}, and this technique
is expected to establish in a simple way the covariant
gauge-independent status of the strong-coupling problem\footnote{
Depending on gauge-fixing procedure for both bulk and brane
diffeomorphisms the diagrammatic technique for this integral can
be essentially simplified by factorizing the bulk and
brane-to-brane parts \cite{QEAB}, which allows one efficiently to
disentangle the different brane and bulk scales of the model.}.

To summarize, we extended the suggestion of \cite{nonloc} for
infrared modifications of Einstein theory to the general case
characterized by two nonlocal operators in Ricci tensor and scalar
quadratic forms and considered restrictions on these operators
imposed by the absence of VDVZ discontinuity. In models with low
strong-coupling scale we showed that their effective action also
contains the strongly coupled mode which cannot be perturbatively
integrated out in terms of the metric variables. It serves in the
lowest order as a Lagrange multiplyer for the constraint on metric
variables which enforces the curvature scalar to vanish on shell.
We demonstrated this situation on the examples of FP, DGP models
and the recently suggested constrained version of the DGP model.

\appendix
\renewcommand{\thesection}{\Alph{section}.}
\renewcommand{\theequation}{\Alph{section}.\arabic{equation}}
\section{Gravitational potentials in the constrained DGP model}
Bulk equations of motion for the action (\ref{10.25}) modified by
the term (\ref{5.1}) read
	\begin{eqnarray}
	&&{\cal E}^{(5)}_{\mu\nu}
	+\frac1\beta\partial_{(\mu}B_{\nu)}=0,  \label{A.5a}\\
	&&{\cal E}^{(5)}_{5\mu}
	+\frac1{2\gamma}\partial_\mu B_5=0,  \label{A.5b}\\
	&&{\cal E}^{(5)}_{55}
	-\frac1\beta\partial_\mu B^\mu=0,      \label{A.5c}
	\end{eqnarray}
where ${\cal E}_{AB}=R_{AB}-\frac12 G_{AB}R^{(5)}$
is the 5-dimensional Einstein tensor. Bianchi identities
$\nabla^B{\cal E}_{AB}^{(5)}=0$ then yield
the equations of motion for gauge conditions
	\begin{eqnarray}
	&&\Box_5 B_5=0,               \label{A.6}\\
	&&B_\mu=-\frac\beta{2\gamma}
	\partial_\mu\frac1\Box\partial_5 B_5,   \label{A.7}
	\end{eqnarray}
in which only $B_5$ is propagating in the bulk, while $B_\mu$ express
algebraically in terms of $B_5$.

From (\ref{A.5b})-(\ref{A.5c}) one gets the $5A$-components of the
metric
	\begin{eqnarray}
	&&H_{5\mu}=\frac{\partial_5}\Box(\partial^\nu
	H_{\mu\nu}-\partial_\mu H)-(\gamma+1)\,
	\partial_\mu\frac{\partial_5}{\Box^2}
	(\partial\partial H-\Box H),            \label{A.8}\\
	&&H_{55}=\frac1\Box\partial\partial H
	-\frac12\beta\frac1\Box
	(\partial\partial H-\Box H),           \label{A.11}
	\end{eqnarray}
whence
	\begin{eqnarray}
	B_5=\partial^\mu H_{5\mu}=
	-\gamma\,\frac{\partial_5}{\Box}
	(\partial\partial H-\Box H),\,\,\,\,
	B_\mu=\frac12\beta\partial_\mu
	\frac{\partial_5^2}{\Box^2}
	(\partial\partial H-\Box H),        \label{A.9}
	\end{eqnarray}
where $\partial\partial H\equiv\partial^\mu\partial^\nu H_{\mu\nu}$
and $H\equiv\eta^{\mu\nu}H_{\mu\nu}$. Combining (\ref{A.7}),
(\ref{A.9}) and the definition of $B_\mu$ one has
	\begin{eqnarray}
	\partial^\mu B_\mu=\Box h_{55}-\partial\partial H
	=\frac\beta2\frac{\partial_5^2}\Box
	(\partial\partial H-\Box H).           \label{A.12}
	\end{eqnarray}
Together with (\ref{A.11}) this implies
	\begin{eqnarray}
	\Box_5(\partial\partial H-\Box H)=0.    \label{A.13}
	\end{eqnarray}
The trace of (\ref{A.5a}) then gives
	\begin{eqnarray}
	\Box_5 H-(2\gamma+\beta/2)
	(\partial\partial H-\Box H)=0.    \label{A.14}
	\end{eqnarray}

With the 4-dimensional transverse-traceless decomposition of $H_{\mu\nu}$
($\eta^{\mu\nu}H_{\mu\nu}^{TT}=0$, $\partial^\nu H_{\mu\nu}^{TT}=0$,
$\partial^\mu\varphi_\mu=0$)
	\begin{eqnarray}
	H_{\mu\nu}= H_{\mu\nu}^{TT}
	+\partial_\mu\varphi_\nu
	+\partial_\nu\varphi_\mu
	+\eta_{\mu\nu}\Phi+
	\partial_\mu\partial_\nu\Psi,    \label{A.15}
	\end{eqnarray}
the gauge becomes $B_\mu=-\Box\varphi_\mu+\partial_\mu(h_{55}
-\Phi-\Box\Psi)$. On the other hand, from (\ref{A.7}) and
(\ref{A.9}) it equals
$B_\mu=3\beta\partial_\mu\varphi/2$. Comparison of these
expressions gives $\varphi_\mu=0$.

Thus, finally from (\ref{A.13}), (\ref{A.14}) and the
transverse-traceless part of (\ref{A.5a}) we have
the set of bulk equations of motion for nonvanishing metric
components
	\begin{eqnarray}
	&&\Box_5 H_{\mu\nu}^{TT}=0,            \label{A.19}\\
	&&\Box_5\Phi =0,                    \label{A.20}\\
	&&\Box_5\Psi
	+3(2\gamma+\beta/2)\Phi=0   \label{A.21}
	\end{eqnarray}
and
	\begin{eqnarray}
	H_{5\mu}=
	3\gamma\frac{\partial_5}\Box
	\partial_\mu\Phi,\,\,\,
	H_{55}=(1+3\beta/2)\Phi
	+\Box\Psi.                            \label{A.23}
	\end{eqnarray}

Linearized Israel junction condition for the action (\ref{10.25}) with
the additional term (\ref{5.1}) (in the presence of conserved matter
sources on the brane) reads
   	\begin{eqnarray}
	m(K_{\mu\nu}-\eta_{\mu\nu}K)
	-{\cal E}^{(4)}_{\mu\nu}
	+\frac1{2\sigma}(\partial_\mu F_\nu
	+\partial_\nu F_\mu
	-\eta_{\mu\nu}\partial_\alpha F^\alpha)+
	\frac{8\pi}{M_4^2}T_{\mu\nu}=0,                   \label{A.24}
	\end{eqnarray}
where $K_{\mu\nu}=(1/2)(\partial_5 H_{\mu\nu}
	-\partial_\mu H_{\nu 5}
	-\partial_\nu H_{\mu 5})$
is the extrinsic curvature of the brane. Transverse-traceless part
of (\ref{A.24}) gives
   	\begin{eqnarray}
	(m\partial_5+\Box)H_{\mu\nu}^{TT}\Big|
	=-\frac{16\pi}{M_4^2} T_{\mu\nu}^{TT},        \label{A.26}
	\end{eqnarray}
while the longitudinal and trace parts imply
   	\begin{eqnarray}
	&&\left(3m\partial_5
	+\frac1\sigma\Box\right)\Phi\,\Big|
	-\frac1{2\sigma}\Box^2\Psi\,\Big|=0,        \label{A.27}\\
	&&\left((9\gamma-6)m\partial_5
	+\Big(\frac1{2\sigma}-3\Big)
	\Box\right)\Phi\,\Big|
	-\frac12\left(3m\partial_5
	+\frac1\sigma\Box\right)
	\Box\Psi\,\Big|+\frac{8\pi}{M_4^2}T=0.      \label{A.28}
	\end{eqnarray}

For $\beta=\gamma=0$ the model essentially simplifies. Equations
(\ref{A.20}) and (\ref{A.21}) decouple and under zero boundary
conditions at $y=\infty$ yield
   	\begin{eqnarray}
	\Phi(x,y)=e^{-y\Delta}\Phi\,\Big|,\,\,\,\,
	\Psi(x,y)=e^{-y\Delta}\Psi\,\Big|,
	\end{eqnarray}
so that everywhere in equations (\ref{A.26})-(\ref{A.28}) $\partial_5$
can be replaced by the (nonlocal) operator on the brane $-\Delta$. Their
solution on the brane becomes
   	\begin{eqnarray}
	\Phi\,\Big|=
	\frac{8\pi}{3M_4^2}\frac1{\Box-
	4m\Delta-3\sigma m^2}\,T, \,\,\,\,
	\Psi\,\Big|=
	2\frac{\Box-3\sigma m\Delta}
	{\Box^2}\,\Phi\,\Big|,       \label{A.32}
	\end{eqnarray}
which finally leads to the gravitational potential
(\ref{33})-(\ref{35}).

Note that the (redundant from the gauge-fixing viewpoint) brane
gauge term in (\ref{A.24}) precludes from enforcing the bulk gauges
and, thus, gives rise to gauge dependence discussed in Sect.5.
Taking the 4-dimensional divergence of the Israel junction condition
(\ref{A.24}) and using the Gauss-Codazzi equation
$\nabla^\mu(K_{\mu\nu}-\eta_{\mu\nu}K)=R_{5\nu}^{(5)}$
one has in virtue of (\ref{A.5b})
	\begin{equation}
	\frac1\gamma\partial_\mu B_5\,\Big|
	=\frac1{m\sigma}\Box F_\mu.         \label{A.33}
	\end{equation}
Thus, the equation for gauge conditions in the bulk (\ref{A.6})
has nonzero data on the brane and has a nonvanishing solution.
Therefore bulk gauge conditions are not enforced, and
their contribution to equations of motion (\ref{A.5a})-(\ref{A.5c})
(proportional to $B_5/\gamma$) is nonvanishing even on shell. This
situation holds also in the limit of $\gamma\to 0$.

\section*{Acknowledgements}

The author would like to thank G.Gabadadze, M.Libanov, V.A.Rubakov
and D.V.Ne\-ste\-rov for helpful stimulating discussions. This work was
supported by the Russian Foundation for Basic Research under the
grant No 05-02-17661 and the LSS grant No 1578.2003.2.

\end{document}